\documentclass[graybox, envcountchap]{svmult}

\usepackage{mathptmx}        
\usepackage{amsmath}
\usepackage{amssymb}
\usepackage{color}
\usepackage{helvet}          
\usepackage{courier}         
\usepackage{dirtree}

\usepackage{makeidx}        
\usepackage{graphicx}        
\usepackage{subfig}

\usepackage{multicol}        
\usepackage[bottom]{footmisc}

\usepackage{hyperref}        
\hypersetup{colorlinks=true,urlcolor=blue}

\usepackage[misc]{ifsym}

\makeindex             

\begin{document}


\title{The Impact of Dust on Cepheid and Type Ia Supernova Distances}
\author{Dillon Brout and Adam Riess}
\institute{Dillon Brout (\Letter) \at Boston University, 685 Commonwealth Ave, Boston, MA 02215, \email{dbrout@bu.edu}
\and Adam Riess \at Space Telescope Science Institute, 3700 San Martin Drive, Baltimore, MD 21218, USA \\ Department of Physics and Astronomy, Johns Hopkins University, Baltimore, MD 21218, USA}

%

%
\maketitle

\abstract{Milky-Way and intergalactic dust extinction and reddening must be accounted for in measurements of distances throughout the universe. This work provides a comprehensive review of the various impacts of cosmic dust focusing specifically on its effects on two key distance indicators used in the distance ladder: Cepheid variable stars and Type Ia supernovae. 
We review the formalism used for computing and accounting for dust extinction and reddening as a function of wavelength. We also detail the current state of the art knowledge of dust properties in the Milky Way and in host galaxies. We discuss how dust has been accounted for in both the Cepheid and SN distance measurements. Finally, we show how current uncertainties on dust modeling impact the inferred luminosities and distances, but that measurements of the Hubble constant remain robust to these uncertainties.}

\section{Introduction}
\label{sec:intro}

Cosmic dust particles play a crucial role in cosmological observations because they absorb, scatter, and re-emit starlight, leading to extinction, attenuation, and an overall reddening of light from distant objects. 
The combination of Milky-Way dust and extragalactic dust affect nearly all measurements in cosmology. From measurements of the cosmic microwave background, to Cepheids and Type Ia Supernovae (SNe Ia), to weak lensing and large-scale structure, to optical counterpart identification of gravitational-wave multimessenger phenomena. In this chapter we focus on the impact of dust specifically for two of the rungs of the SH0ES \cite{Riess2022} cosmic distance ladder: Cepheids and SNe~Ia. 

Cosmic dust grains are composed of tiny solid particles, ranging in size from a few nanometers to several micrometers, that are generated by a variety of processes, including stellar nucleosynthesis, supernova explosions, and collisions between solid bodies. While the structure and composition of dust particles depend on the specific astrophysical environment in which they are formed, in general, dust grains are made up of silicon, carbon, and large hydrocarbon molecules (See Figure \ref{fig:extinction} Top), each of which have slightly different effects due to their composition, shape, and reflectivity. The particles can have a wide range of shapes, from spherical to elongated, though their shape and size distributions are not measured directly, but rather are instead inferred through their effects on extinction, polarization, and albedo.

To measure for dust extinction in the Milky Way, specific backlights such as stars and quasars serve as reference sources with well-characterized intrinsic spectral energy distributions (SEDs). By comparing the observed attenuated SEDs to the unattenuated counterparts, one can estimate the extinction as a function of wavelength. In this chapter, we will explore the mathematical relationships between relevant parameters describing the physics and impact of cosmic dust, including extinction, reddening, and dust laws, with a particular focus on their impact on the cosmic distance ladder and measurements of cosmological parameters.

\section{Dust formalism}
\label{sec:formalism}
Dust extinction and attenuation are processes that cause dimming and reddening of light. While they arise from different but related physical mechanisms, they have the same effects on cosmological observables. \textbf{Extinction} is due to dust along the line of sight to objects that causes absorption or scattering away from the line of sight.
\textbf{Attenuation} includes the effects of extinction but also includes scattering back into the line of sight and the contribution to the light by unobscured stars. It is the net loss of light from a complex, spatially extended population and depends on both extinction and the star-dust environment geometry. The wavelength dependent extinction or attenuation can be expressed mathematically as:
\begin{eqnarray}
A_\lambda = m_\lambda – m_{\lambda_0}
\label{eq:extinction}
\end{eqnarray}
where $m_{\lambda_0}$ represents the unextinguished magnitude of an astronomical object at a given wavelength, $m_\lambda$ represents the observed magnitude, and $A_\lambda$ describes the amount of extinction. 

\begin{figure}
\centering\includegraphics[scale=.37]{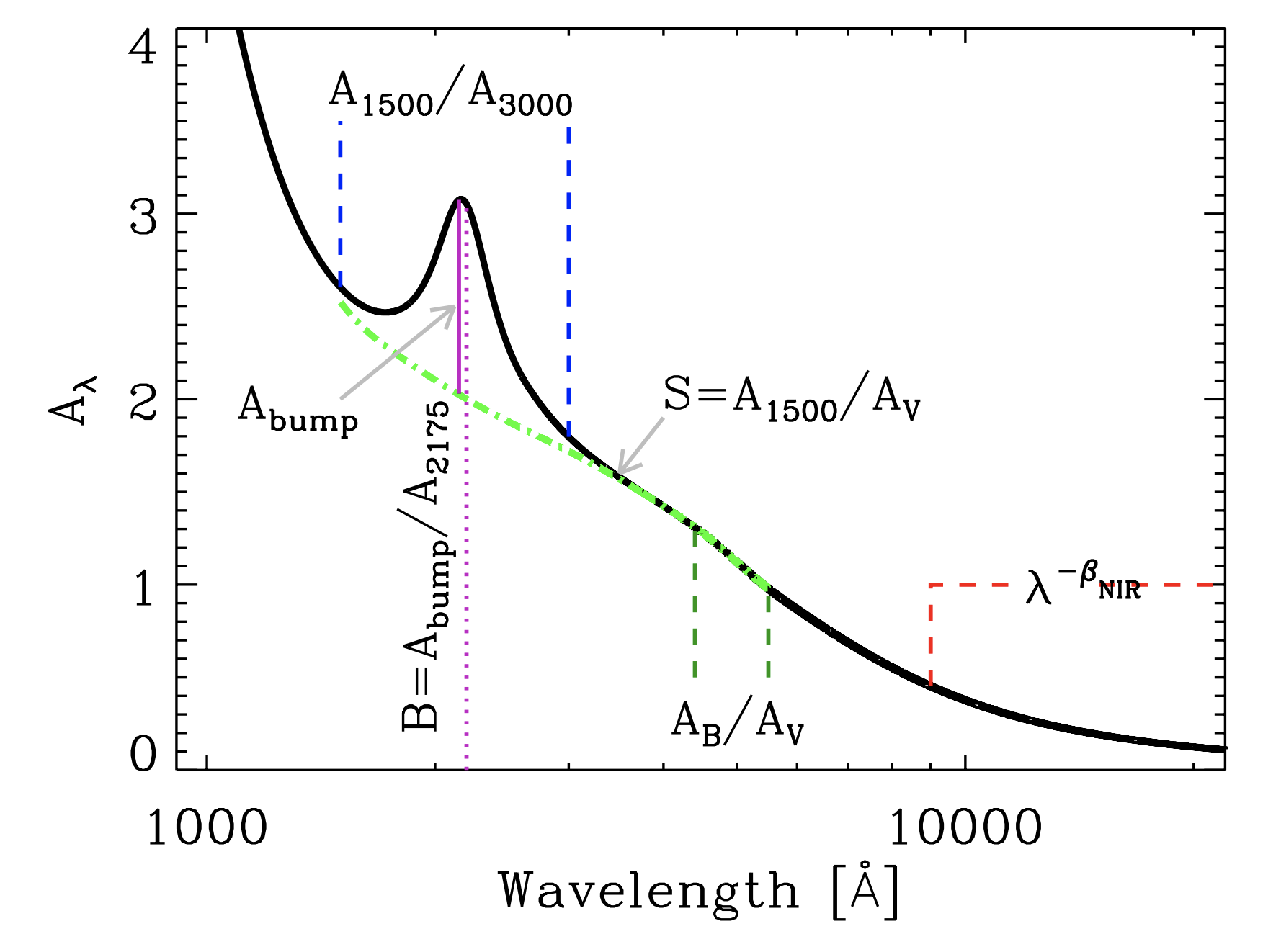}
\centering\includegraphics[scale=.48]{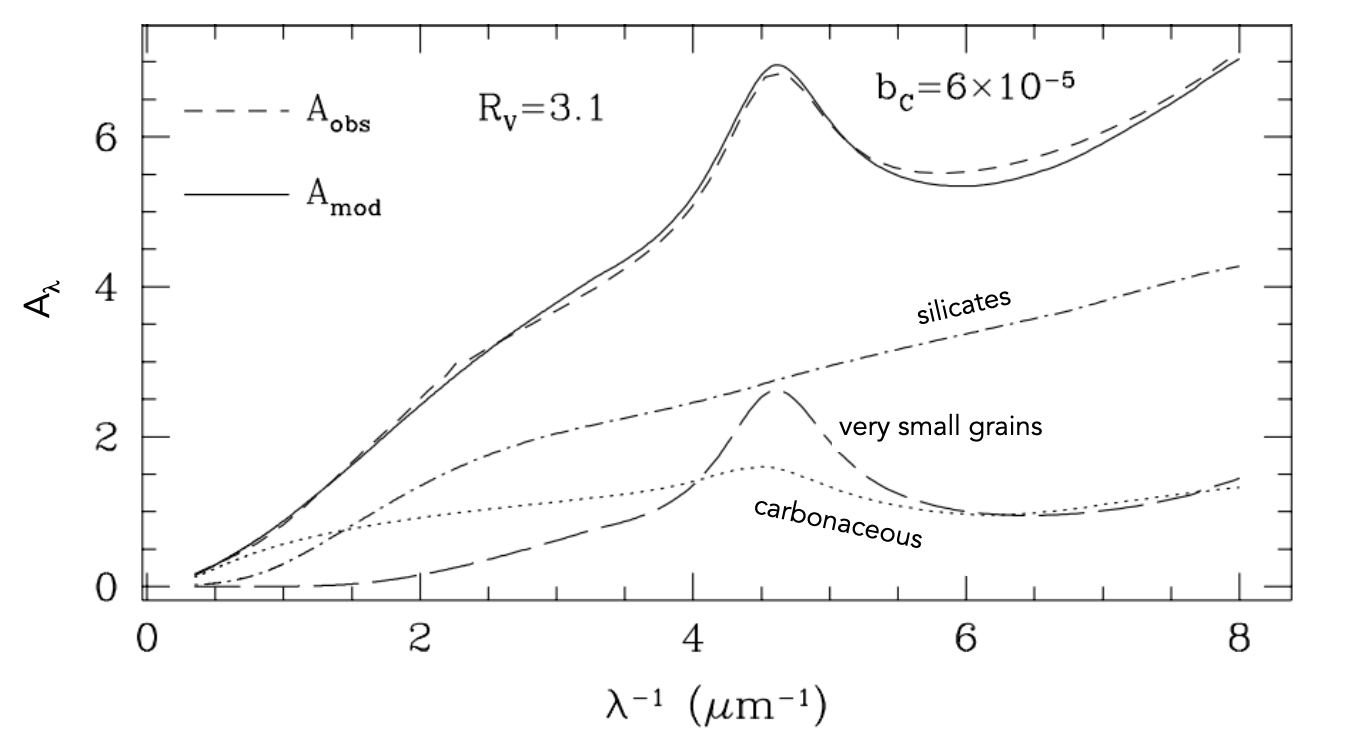}
%
%
\caption{\textbf{Top:} Schematic representation \cite{Salim_2020} of an example non-normalized attenuation or extinction curve.
\textbf{Bottom:} Schematic representation \cite{Weingartner_2001} of the different effects of grain compositions.}
\label{fig:extinction}       
\end{figure}

The attenuation or extinction \textit{curve} describes a normalized shape of $A_\lambda$ as a function of wavelength (See Figure~\ref{fig:extinction}~Bottom). The overall normalization (in most cases set by $A_V$) of the attenuation or extinction curve is set by the column density of grains and the overall shape is set by the mixture of grain types and grain size distribution, and grain optical properties along the line of sight.  
 It is difficult to relate the extinction curve directly to dust physics, so such curves are largely a measured quantity using known background sources and for the purposes of cosmology this is sufficient.
 
There are several observed features of a typical dust extinction curve (shown in Figure~\ref{fig:extinction}). The Ultra-violet (UV) - optical slope ($A_{\rm 1500}/A_{\rm 3000}$) and the strength of the UV bump (B) is seen to vary significantly within and across galaxies and provides a large lever arm to constrain the overall dust properties as it provides information about the ratio of the star formation rate (SFR) and the dust column density. The optical slope ($A_B/A_V$) is important to account for in many cosmological constraints that are observed in the optical, e.g. SNe Ia. The near-IR power-law slope ($\beta_{\rm NIR}$) which is the smallest of these features as dust has the least impact and variation at larger wavelengths.

The color excess $E(B-V)$ characterizes the reddening of an object's spectrum due to dust extinction. It is defined (Eq. \ref{eq:reddening}) as the difference between the extinction in two bands (B-V). 
\begin{eqnarray}
E(B-V) = A_B – A_V
\label{eq:reddening}
\end{eqnarray}
By comparing the magnitudes in different bands, one can estimate the amount of reddening caused by dust and gain insights into the properties of the intervening material. Differential measurements of E(B-V) are extremely useful, though to know the true color excess one must have some knowledge of the intrinsic color of the source object. Thus, differential maps of E(B-V) across the Milky-Way galaxy are quite accurate, though knowing the absolute value/normalization of the color excess is a more difficult task (see Section X).

Lastly, a useful quantity that is often used in analyses of distance indicators is the `total-to-selective extinction ratio' $R$: 
\begin{eqnarray}
R_V \equiv A_V / E(B-V)
\label{eq:rv}
\end{eqnarray}
where this represents the linear slope of the extinction curve at a given band (e.g. $R_V$). 
These ratios provide quantitative information about the differential extinction in the relevant ($V$-optical) bands/wavelengths and are used in cosmological constraints of Sections~\ref{sec:cepheids}~and~\ref{sec:SN}.

\section{Milky Way Dust}
\label{sec:mw}
Accurate estimates of the Milky Way dust are exceptionally important in cosmology as all electromagnetic observations must contend with it at some level to correct for the observed brightnesses and colors of extra-galactic sources and to understand spatially dependent selection effects. 
\\
\vspace{.1in}

\textbf{Reddening:}
Dust maps can be derived in two main ways: emission-based maps and extinction-based maps. Emission-based maps utilize the observed far-infrared thermal emission from dust to estimate the dust column density. The Schlegel-Finkbeiner-Davis map (SFD: \cite{sfd}) is based on 100$\mu$m data from the Infrared Astronomy Satellite (IRAS), with corrections to remove zodiacal light and the cosmic IR background based on maps from the Diffuse Infrared Background Experiment (DIRBE).  This SFD 100$\mu$m map has a resolution of 6.1 arcminutes FWHM. Because hotter dust emits more 100$\mu$m radiation per column density, the SFD 100$\mu$m map is modulated by a temperature correction based on the DIRBE 100$\mu$m/240$\mu$m color temperature. This modulation is on a 1.3 degree scale where the dust is sufficiently bright, and transitions to a 7 degree scale at high Galactic latitude. 

The overall calibration of the SFD map was updated in 2011 \cite{schlafly11} based on spectra of main sequence turn-off stars. This followed \cite{schlafly10}, which used Sloan Digital Sky Survey data of the ``blue tip" of the stellar locus (a prominent blue edge in the stellar color distribution) that can be used to estimate extinction along the line of sight and is sensitive to $\sim10-20$mmag color changes. Using this method \cite{schlafly10} find a 14\% scaling of the overall SFD map normalization and a preference for the Fitzpatrick 1999 \cite{f99} color law over O'Donnell (1994) \cite{o94} and Cardelli et al. (1989) \cite{ccm} (see next section). \cite{schlafly11} follow up on this work using a different technique that uses stellar spectra and modeling codes to find an additional 4\% calibration normalization relative to \cite{schlafly10} and re-confirm a preference for the Fitzpatrick color law in the optical wavelengths. This 4\% difference between modern methodologies has been adopted for the systematic uncertainty in numerous cosmology analyses. Though it is important to emphasize that these uncertainty estimates were best constrained from data within the galactic plane and are not guaranteed to hold true out of the plane of the galaxy where data is more sparse and relative constraints are worse (though with lower reddening amplitude).

After the SFD work, maps based on Planck \cite{a16} became available.  Planck observations of dust at several frequencies (217, 353, 545, 857 GHz) are used, but in order to constrain the high-frequency side of the thermal emission spectrum, they used the SFD 100$\mu$m.  In other words, the same deficiencies in the SFD zodiacal light correction and zero point affect the Planck maps.  Furthermore, the Planck analysis attempts to infer the emissivity power-law index $\beta$ along with dust temperature and optical depth $\tau$, introducing additional noise at high latitude where the emission is to faint to constrain 3 parameters.  So although the Planck data are superior to IRAS in general, the derived dust map shares some shortcomings with SFD. 

Another method for mapping dust involves utilizing inferred stellar extinctions while considering a presumed or fixed reddening law. Empirical constraints on extinction require some knowledge of the unextinguished spectrum, either via models or empirical pairs of more/less reddened stars. Estimations of reddening along the line of sight to these stars is somewhat challenging, but not an overwhelming burden.
Furthermore, since each star provides information about the extinction up to its distance, these stellar reddening-based maps allow for the reconstruction of dust extinction in three dimensions \cite{Green3d,Vergely_2022,edenhofer2023parsecscale}. However, these maps may suffer from contamination due to incomplete removal of galaxies and quasi-stellar objects (QSOs) from stellar catalogs \cite{Chiang_2019}. This contamination can lead to systematic errors, causing both overestimation and underestimation of extinction at the positions of galaxies and QSOs. Thus, these maps are not perfect and there is the need to motivate further progress for the next generation of experiments. 
\\
\vspace{.1in}

\textbf{Dust Laws:}
For use in broadband photometry fits like that of SN Ia and Cepheid luminosity measurements, it is important that attenuation curves are parameterized for use in fitting. Common parameterizations distill attenuation to single parameter characterizing the slope of of a predefined curve, though some include additional parameters describing the UV bump (see Fig~\ref{fig:extinction}). Such parameterizations do not need to be unique to the Milky Way galaxy and can extend to extra-galactic dust measurements for comparison. While the variations in attenuation curves has thus far been difficult to relate to dust physics, for the purposes of precision cosmological constraints the empirical relations described below are adequate. 

Fitzpatrick (1999) \cite{f99} provide a single-family parameterization for extinction curves (originally intended for the Milky-Way, though now extended beyond) from the optical to near-IR. They used a comprehensive dataset of reddened stars with known intrinsic colors and distances, coupled with detailed spectroscopic measurements to compute reddening effects across a wide range of wavelengths and characterize R(V). Fitzpatrick (1999) \cite{f99} showed significant variations in the extinction law along different sight lines and several notable studies have extensively confirmed and characterized these variations (\cite{Draine_2003,schlafly10,schlafly11}). The impact of this variation is discussed in \ref{sec:SN}

\section{Extragalactic Dust}
\label{sec:dustvar}

For many cosmic distance probes, it is also important to understand the dust in the galaxy to which distances are being measured. As $R_V$ varies for different dust grain sizes and composition, and galaxies have different dust properties, it is well known that different galaxies and different regions within galaxies exhibit a wide range of $R_V$ values. In fact, while the Milky Way galaxy has an $R_V$ on average $\sim3.1$, it has a distribution of at least $\sigma_{R_V}=0.2$ \cite{Schlafly16}. Additionally, different parts of the LMC and SMC have been found to have $R_V$ values with a range of $R_V\sim2-5$ \cite{Gao2013,Yanchulova17}. The origin of these dramatically different extinction curves is still debated as it could be driven by metallicity, radiation field strength, or both.

Furthermore, \cite{Salim18} study the dust attenuation curves of 230,000 individual galaxies in the local universe, using GALEX, SDSS, and WISE photometry calibrated on the Herschel ATLAS, and they find quiescent galaxies, which are typically high-mass, have a mean $R_V=2.61$ and star-forming galaxies, which are lower-mass on average, have a mean $R_V=3.15$. 

$R_V$ has also been measured through large SN sample statistics and detailed studies of individual SNe, though often with varying sets of assumptions. \cite{Cikota16} compiled 13 various studies of SN~Ia samples from the literature which determined a range of $R_V$ values from $\sim1$ to $\sim 3.5$. \cite{Cikota16} itself determined $R_V$ from nearby SNe and for 21 SNe Ia observed in Sab-Sbp galaxies and 34 SNe in Sbc-Scp they find $R_V=2.71\pm1.58$ and $R_V = 1.70 \pm 0.38$ respectively. While so many past analyses have recovered $R_V<2$ for studies of individual SNe (e.g. \cite{XWang05,Krisciunas06}), these were often SNe~Ia with high $E(B-V)$, and it was postulated $R_V$ may decrease with $E(B-V)$. However, \cite{Nobili_2008} found from a sample of modestly reddened ($E(B-V)< 0.25$ mag) SNe~Ia, a small value of $R_V\sim1$ and more recently, \cite{Amanullah15} analyzed high-quality UV-NIR spectra of 6 SNe and found that SNe with high reddening indicated $R_V$'s ranging from $\sim1.4$ to $\sim2.8$ and SNe with low amounts of reddening also indicated $R_V$'s of $\sim1.4$ and $\sim2.8$. Importantly, \cite{Amanullah15} stressed that the observed diversity in $R_V$ is not accounted for in analyses that measure the cosmological expansion of the universe.

Since the low $R_V$ values ($<2$) are not found in studies of the Milky Way, this has motivated various SN~Ia studies to ascribe the dust to circumstellar dust around the progenitor at the time of the explosion \cite{Wang05,Goobar08}. However, an alternative interpretation could be that the low $R_V$ values are caused by dust in the interstellar medium \cite{Phillips13}. This understanding has been supported by \cite{Bulla18a,Bulla18b}, which constrained the location of the dust that caused the reddening in the SN~Ia spectra to be, for the majority of the SNe that they observed, on scales of the interstellar medium, rather than circumstellar surroundings.

\section{Impact on Distance Indicators}
\label{sec:distances}

\subsection{Impact on Cepheid Magnitudes}
\label{sec:cepheids}

While the impact of dust on measurements of galactic and extra-galactic Cepheids is minimal due to the fact that in the distance ladder approaches they are observed in the Hubble Space Telescope (HST) Near Infrared (NIR) bands, we detail here the possible impact and its relevance for the inference of the Hubble Constant. For the case of nearby first and second rung Cepheids in the SH0ES distance ladder, they are all at extremely low redshift ($z<0.01$), and therefore both Milky Way and extra-galactic dust affect the same range of wavelength and their effects are treated simultaneously below. 

Extra-galactic Cepheids are observed in regions of recent
star formation in late-type galaxies and are found to have a mean reddening of $E(V-I)$ $\sim0.3$mag. The SH0ES \cite{Riess2022} observations in HST NIR F160W ($H$-Band) 
reduces the impact of and correct (`deredden') for the effects of interstellar dust reddening. 

For the distance ladder, the goal is to measure the relative distance between two sets of Cepheids (rungs) free from reddening which by definition is $(m_{H,i}-A_{H,i})-(m_{H,j}-A_{H,j})$ for the $ith$ and $jth$ samples where $A_H$ is the extinction from dust and comes from the color excess, $E(V-I)$ multiplied by a reddening ratio, $R_H$, so $A_H=R_H \times E(V-I)$.  However, in practice we do not directly measure $E(V-I)$ which would require knowledge of the intrinsic color, $(m_V-m_I)_0$. Rather, we observe the apparent color, $(m_V-m_I)_0$.  A useful, shortcut to measuring the distance difference using only the apparent quantities is to use so-called ``Wesenheit"  magnitudes \cite{madore82} ($m_H^W$) following: 

\begin{eqnarray}
m_H^W = m_H – R_H \times (m_V-m_I)
\label{eq:dereddened}
\end{eqnarray}

This measure is based only on directly observed quantities.  Below we denote the components of the Cepheid model that make up $(m_V-m_I)$.
Wesenheit magnitudes $m_H^W$ are not true magnitudes in the sense that they do not represent the brightness.  Rather, their utility is in the comparison between two sets of such magnitudes, which results in the reddening-free magnitude difference.  

Their utility can be seen when we decompose the apparent color $(m_V-m_I)$ into its constituents.  The main component is $(m_V-m_I)_0$ which would be the ``unreddened'' color at the middle of the Cepheid instability strip (typically this term is $V-I \sim 1$).  The displacement in color from the midline of the instability strip is given by $\Delta (m_V-m_I)$ which corresponds to an intrinsic property of each Cepheid. Finally, the color excess due to dust is given by $E(V-I)$ which is reddening by dust:

\begin{eqnarray}
m_V-m_I =(m_V-m_I)_0 + \Delta_{V,I} + E(V-I)
\label{eq:colorexcess}
\end{eqnarray}
 
 Thus, multiplying ($m_V-m_I$) of Eq.~\ref{eq:dereddened} by a single $R_H$ for two samples $i$ and $j$ (i.e. two rungs of the distance ladder) results in the following when applied to the decomposition in ~\ref{eq:colorexcess}. The terms $(m_V-m_I)_0$ multiplied by $R_H$ will be the same for different galaxies, this being an intrinsic property of Cepheids, and it is important to use the same value of $R_H$ (whatever that is) where multiplying by the intrinsic color. 
When multiplying $R_H$ on the $\Delta_{V,I}$ term, because the mean of $\Delta_{V,I}$ is at or near zero (by definition) for large samples of Cepheids that fill the instability strip, $R_H\Delta_{V,I}$ will also average to zero for a statistically large sample.  However, because Cepheids, like most stars, are dimmer when they are redder (cooler), a positive value of $R_H$ can also reduce the apparent dispersion of the instability strip.  In fact, observations of stars show the change in intrinsic color with brightness is of the same order as the reddening law, $R$, a happy coincidence, so that the term $R_H$ reduces variance in two ways.  Therefore, it is common to choose the value of $R_H$ to be the reddening ratio between the $H$-band and the $V-I$ color, as derived from a reddening law. Lastly, the only term that does not cancel or we expect to not be consistent across galaxies is $R_H\times E(V-I)$ because $E(V-I)$ varies within and across Cepheid samples. 
Thus, we can now rewrite the observable of Eq.~\ref{eq:dereddened} as $m_H-R_H\times E(V-I)$. Finally, taking the differential between two samples leaves us with our desired result $m^W_{H,i} - m^W_{H,j} = (m_{H,i} - R_H\times E(V-I)_i) - (m_{H,j} - R_H\times E(V-I)_j)$.



We get the exact difference in distance when $R_{W,i}=R_{W,j}$ (and with a large Cepheid sample, $\Delta_{V,I} = 0$).  We note here that the ``shortcut'' equation~\ref{eq:dereddened} only works if we require $R_{W,i}=R_{W,j}$. If we do not want to require $R_{W,i}=R_{W,j}$, then we must decompose \ref{eq:dereddened} into the terms in \ref{eq:colorexcess} and then we need to define and model the intrinsic standardizable color of Cepheids from another sample and only allow for $R_W$ to vary across for the dust excess component \cite{Follin_2018,mortsell,Riess2022}.

In the case of correcting NIR magnitudes ($m_H$) following Eq.~\ref{eq:dereddened}, we have thus far assumed a single $R_H$. Fortunately, $R_H$ is well-characterized to be small ($\sim0.4$) for Milky Way-like extinction. This value is much smaller than for correcting optical magnitudes ($R_V\sim3$) and demonstrates the comparatively low sensitivity to dust in the SH0ES analysis. This dust law insensitivity is also important when considering potential variation in the dust curve across the population of Cepheid host galaxies. While $R_V$ can range from 2-6, $R_H$ correspondingly ranges from 0.35-0.5 with a definitive lower bound at $\sim$ 0.2 due to the Rayleigh scattering limit or 0.25 for any known reddening law. 

While $R_H$ has been determined independently, $R_H$ can also be determined from the Cepheids themselves as the value that minimizes the observed scatter in the observed Period-Luminosity relation. From the highest quality Milky Way Cepheids $R_H=0.363\pm0.038$ \cite{Riess_2021} which matches independently determined values in the literature. Additionally, $R_H$ determined from the entire distance ladder of Cepheids is $R_H=0.34\pm0.02$ \cite{Riess2022}. For this value, $H_0$ increases by 0.3. 

It is possible to consider different values of the reddening ratio for each individual host.  However, because with different $R_H$ the intrinsic color term $(m_V-m_I)_0$ no longer cancels across rungs, the intrinsic term loses its meaning, and defies the interpretation of Cepheids as standard candles. So this cannot be done directly with Wesenheit magnitudes. Therefore, to vary the reddening ratio, one must first subtract the color excess using an empirical period-color relation, such as $ (V-I)_0 = 0.5+0.25 \log P$ (empirical, Milky Way or Large Magellanic Cloud) from $(V-I)$, or this can lead to a spurious result \cite{mortsell}. When $R_H$ is fit individually for each host, $H_0$ has been shown to change at the 0.8 km/s/Mpc level, as shown in \cite{Follin_2018}. In practice, the Cepheid data is not very constraining of individual reddening ratios in the SN Ia hosts, so an alternative and better approach is to assign individual host reddening ratios based on the host star formation rate and mass which predict $2.8<R_V<4.2$ \cite{Riess2022}. 

Observing in the NIR to mitigate the impact of dust is not without its own challenges. On HST, the resolution in NIR is 2-3 times worse than that of the optical and in the NIR the background of ubiquitous red giant stars is an order of magnitude higher. However, with the advent of the James Webb Space Telescope (JWST), we have been further reassured that there is not a confounding issue when it comes to the impact of dust for Cepheids. Both \cite{Yuan_2022} and \cite{riess2023crowded} show that JWST provides significantly improved resolution, improved sensitivity in the NIR, and even lower impact of interstellar dust in comparison to HST. They find good agreement between the HST and JWST systems and no evidence for bias beyond the 0.04mag level, which is significantly better than the 0.2mag size of the $H_0$ tension and is expected to further improve with future observations.

\subsection{Impact on Type Ia Supernova Distance Moduli}
\label{sec:SN}

In order to compute distances, Type Ia Supernova (SN Ia) photometric light curves are fit to an underlying spectral time-series model (e.g. SALT3 \cite{SALT3}). These fits output the following summary statistics that characterize each light curve: the peak B-band brightness ($m_B$), the duration `stretch' ($x_1$), and the color ($c$). The latter of which, $c$, characterizes the combination of intrinsic SN Ia color and dust reddening because at the photometric light curve level such properties are degenerate. SN Ia light-curve fit parameters are then combined empirically to compute distances ($\mu_\mathrm{obs}$) where the nuisance parameters ($\alpha$, $\beta$, $\gamma$) are single numbers that correlate light-curve and host properties with an intrinsic luminosity and are determined such that they minimize the scatter in distances of SNe Ia \cite{Marriner_2011}. The distance moduli are computed following a modified version of the Tripp \cite{Tripp98} equation:

\begin{equation}
    \mu_\mathrm{obs} = m_B + \alpha x_1 - \beta c + \gamma G_{\rm host} - \mathcal{M}_B + \mu_\mathrm{bias},
    \label{eq:tripp}
\end{equation}
where $G_{\rm host}$ is each host-galaxy's mass that has been shown to correlate with SN Ia brightness \cite{Hicken09a,Sullivan2010,Lampeitl2010,Childress2013}, $\mu_\mathrm{bias}$ is a correction for instrumental and survey selection effects for each SN based on careful survey simulations \cite{Kessler18}, and $\mathcal{M}_B$ is a single number representing the intrinsic magnitude of a SN Ia as determined by the SH0ES distance ladder.

\begin{figure}
\centering\includegraphics[scale=.75]{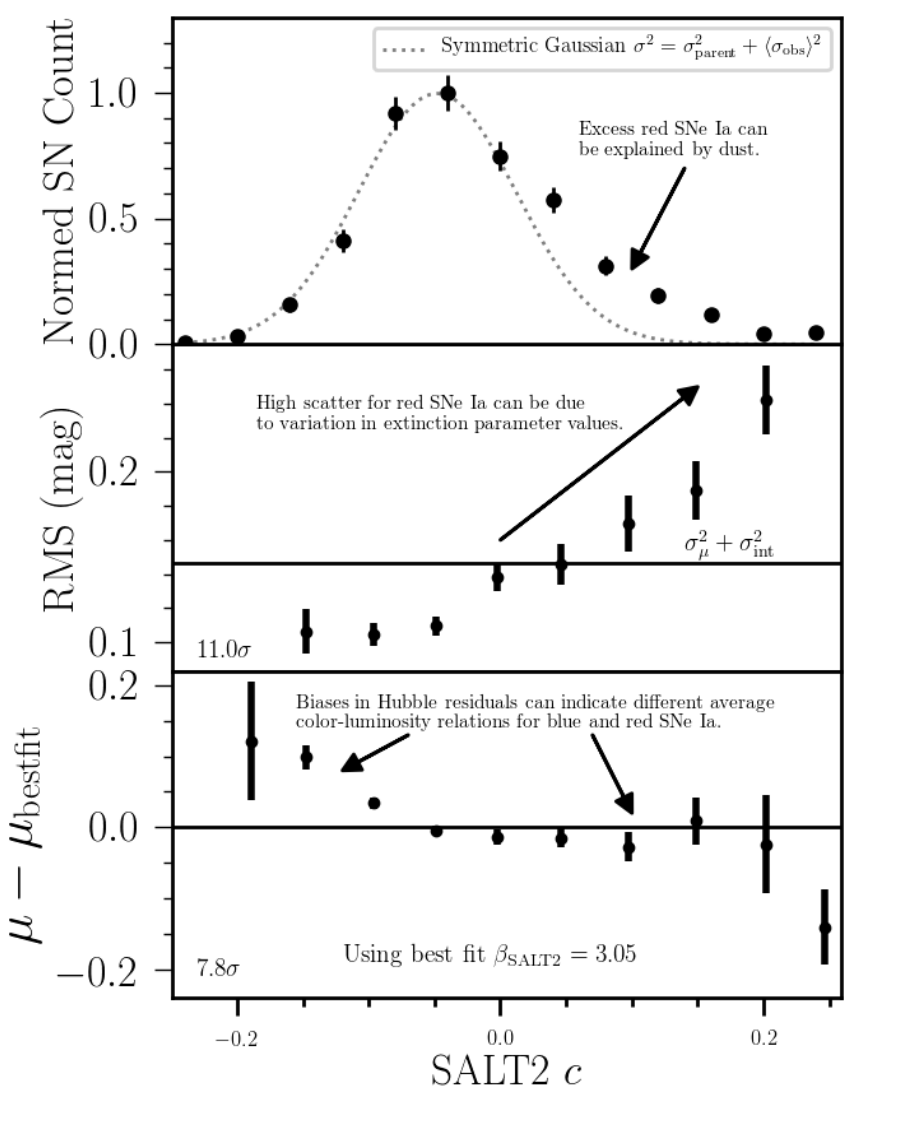}
\caption{\textbf{Top:} Observed SALT light curve fit parameter color ($c$) from the full data sample of \cite{bs21}, with a symmetric Gaussian overlaid (dashed) for comparison. Skewness in the observed dataset to the red is evident and indicative of dust. \textbf{Middle:} RMS scatter of Hubble Diagram residuals as a function of color. The RMS is calculated after Tripp standardization and after subtracting the mean Hubble residual bias. SNe Ia with bluer colors exhibit significantly better standardizability than their redder SN Ia counterparts. This is indicative of unmodeled variation in dust properties for each SN Ia or host galaxy. \textbf{Bottom:} Binned Hubble diagram residuals as a function of color, after Tripp standardization using a single best fit $\beta$. The bluest SNe Ia exhibit a different color-luminosity relation than the reddest; indicative of an intrinsic SN Ia distribution and a secondary distribution characterized by dust. In the bottom two panels, the significance of the deviation from a flat line is show in the bottom corner.}
\label{fig:bs21}       
\end{figure}

For SNe Ia, galactic and extra-galactic dust both reddens and obscures light in the rest-frame UV to optical wavelengths. With the exception of the future NASA Roman Space telescope, large area supernova surveys are typically observed from the ground in the optical wavelengths (due to atmospheric transmission). The analysis of Pantheon+ \cite{Brout2022} and SH0ES \cite{Riess2022} account for Milky Way galactic dust by applying the Schlafly et al. 2011 stellar locus map \cite{schlafly11}. A second correction is applied with a 4\% absolute scale offset following \cite{schlafly10} found using the stellar spectral modeling approach. This difference is evaluated in the Pantheon+ SN Ia covariance matrix in order to characterize the systematic uncertainty. 

Dust in the host galaxy of the supernova is degenerate with the light-curve color and is hypothesized to be the major cause of the so called intrinsic scatter of standardized SN~Ia brightnesses (the excess scatter of SN~Ia distance residuals compared to a best-fit cosmology after accounting for measurement noise). Recent promising models have now explained the majority of SN~Ia intrinsic scatter using dust. The size of SN~Ia intrinsic scatter has recently been found to depend on SN Ia color and host galaxy properties (BS21: \cite{bs21}). Recently, BS21 introduced dust extinction to the SN Ia spectral model by forward modeling simultaneously 1) an intrinsic SN color population, 2) extrinsic dust populations, 3) variation in the dust properties from galaxy to galaxy, and 4) correlations between the dust properties of galaxies with different host properties ($G_{\rm host}$). BS21 marks a major improvement over previous intrinsic scatter models commonly used in SN Ia cosmology analyses (e.g. grey scatter: Guy et al. 2010 \cite{Guy2010} and spectral scatter: Conley 2011 \cite{Conley2010}. While the Tripp equation \ref{eq:tripp} does not disentangle intrinsic color and extrinsic dust properties, by forward modeling this process and analyzing the forward modeled simulations with the Tripp equation, the model can be compared with and fit to the data in a hierarchical approach (see Fig.~\ref{fig:bs21}). 

The major successes of the BS21 model are that solves two key problems with one simple solution. 1) It allows for $R_V$'s in high and low mass host galaxies which explains the observed correlation with host properties. 2) It explains the observed color dependent intrinsic scatter with variation in dust properties. It also makes key predictions, that the typical SN Ia dust (whether it be local to the SN or global to the host galaxy) corresponds to an $R_V\sigma2$ and with a large inter-galaxy dispersion $\sim1$ (BS21 and \cite{popovic21b}) which has similarly been seen in numerous followup studies \cite{Chen_2022,Meldorf_2022,Thorp_2022}.
 Continuing upon the improved modeling developed in BS21, \cite{popovic21a} implement the dust model used in cosmological analyses. \cite{popovic21a} correct the Pantheon+ distance moduli on a population level and \cite{popovic21b} address the computational limitations in the hierarchical model fitting process in order to provide robust model uncertainties that include correlations in uncertainties between dust parameters and SN intrinsic parameters. 

Note that alternative models for the correlation between SN Ia luminosity and host-galaxy properties have been proposed that involve progenitor differences \cite{Rigault2013} or different subclasses \cite{Polin_2019} instead of dust. However, in light of recent findings in Pantheon+ for the very strong color dependence of intrinsic scatter and $\beta$, such alternative scenarios must explain now a three path mechanism to explain the observations of SN Ia color skewness to the red, SN Ia one-sided color-dependant scatter, and both scatter and $\beta$ dependencies on host properties (not depicted in Fig.~\ref{fig:bs21}). The large statistics of the Pantheon+ sample illuminating these dependencies makes such alternative approaches challenging. 

A new SN-host correlations model was recently explored in the Dark Energy Survey Supernova Program that accurately and realistically models physically motivated relations between SN age, SN host galaxy properties, and SN stretch ($x_1$). Vincenzi et al. (in prep) find that this model does not explain the host-mass correlation with luminosity, so they use this model in conjunction with a BS21 dust based scatter model and find that inferred distances to SN Ia remain relatively unchanged relative to the approach taken in Pantheon+. Finally, further evidence for BS21 is substantiated by the fact that dust properties are also found to correlate with other physical parameters of interest in galaxy surveys. \cite{Salim18} find lower mass galaxies tend to have steeper attenuation curves ($R_V$).

\section{Impact on Cosmological Constraints}
\label{sec:cosmology}

\subsection{ Dust Impact (or lack-thereof) on the H0 Tension}
The SH0ES methodology of obtaining the Hubble constant is one that leverages differential measurements between the first and second rungs (for Cepheids), and the second and third rungs (for SNe Ia). Here we emphasized that only differences in dust properties between rungs is what could potentially contribute to biases in H$_0$. This is because the SH0ES likelihood and methodology is a differential measurement for Cepheids and SNe Ia, meaning that offsets that are consistent across rungs cancel.

\vspace{.1in}

\noindent\underline{\textbf{Cepheids:}} 

\begin{figure}[t]
\centering\includegraphics[scale=.4]{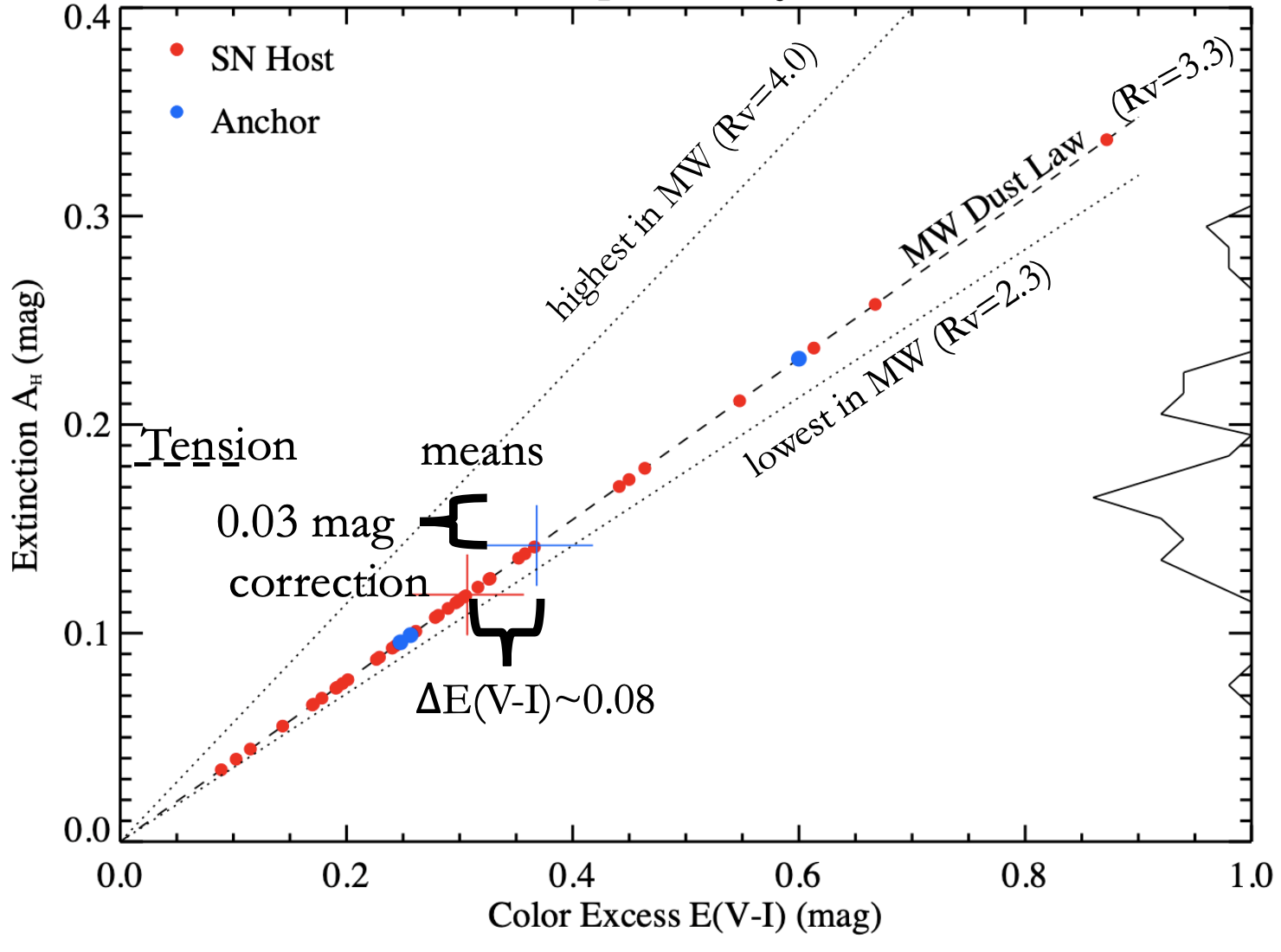}
\caption{Difference between the observed color excess  of the Cepheids for the first (Anchor) and second (SN Host) rungs of the distance ladder for different underlying dust law ($R_V$). $H_0$ Tension of 0.2 mag is shown for comparison to the 0.03 mag differences observed in the sample.}
\label{fig:ladderdust}       
\end{figure}

Figure \ref{fig:ladderdust} visualizes the impact of dust on the Cepheids in the first two rungs of the distance ladder (Geometric ``Anchor" Cepheids and ``SN Host" Cepheids). For the Cepheids in both rungs (red and blue points), color excess $E(V-I)$ is measured relative to the intrinsic Cepheid color at the center of the instability strip and for a typical Milky Way dust law ($R_V=3.3$) the corresponding extinction ($A_V$) values are inferred from the color excess. As discussed in Section~\ref{sec:dustvar}, there are numerous extinction laws not only within a galaxy but also across galaxies. Therefore, Figure~\ref{fig:ladderdust} also visualizes other possible dust laws ranging from $R_V$ of 2-4. 

The difference in mean color excess and inferred extinction is the most relevant quantity for assessing the impact on the $H_0$ tension.  Thus, we show the Cepheid color excess for each rung (First: blue, Second: red) and the average difference ($\Delta E(V-I)\sim0.08$) in Figure \ref{fig:ladderdust}. If the means between rungs are identical, $H_0$ is completely insensitive to dust by nature of the distance ladder methodology, as only differences in the means of each rung can propagate to changes in $H_0$. 

The entire $H_0$ tension between SH0ES and Planck (6km/s/Mpc) can be expressed as 0.18mag in $\Delta A_V$ (shown for reference on Figure~\ref{fig:ladderdust}). However, because the mean color excess between the first and second rungs is $\Delta E(V-I)\sim0.08$ and because a Milky Way dust law of $R_V$=3.3 corresponds to an $R_H$=0.4 in the infrared (because the Cepheid luminosities are computed in HST H-band), this corresponds to a 0.03 $A_H$ difference for `Milky Way-like' dust. If the typical dust law is not Milky Way-Like, but rather $R_V$=2.3 ($R_H=0.35$) or in a worse case scenario of $R_V$=4.0 ($R_H=0.35$), the difference in extinction $A_H$ could range from 0.025 to 0.04, however, in no case does the expected $A_H$ approach the 0.18mag needed to resolve the $H_0$ tension. Additionally, analyses such as \cite{Follin_2018} and \cite{mortsell} have allowed for $R_H$ to vary from galaxy to galaxy and find that, despite the extra model freedom and parameters, because $R_H$ is small and the color excess are small,  variations in $R_H$ across Cepheid galaxies cannot explain the $H_0$ tension. Finally, in Table\ref{tab:variants} are depicted the SH0ES variants; SH0ES explore numerous attenuation laws (e.g. Cardelli, Clayton, \& Mathis (1989) \cite{ccm} and Fitzpatrick (1999) \cite{f99}), attenuation law slopes and dependencies, and even remove the dust treatment altogether and show that they all have either minimal impact on $H_0$ or move the central value up to make the tension worse.

\vspace{.1in}

\noindent\underline{\textbf{SNe Ia:}}  

As shown in Figure 13 of Pantheon+, the impact of a 4\% absolute scaling of galactic dust maps on $H_0$ is negligible. This is because, first, SNe Ia are typically discovered in low extinction regions of the sky. Second, the average Milky Way $E(B-V)$ of the second rung is 0.050 and of the third rung is 0.046; multiplying the difference of 0.004 by the typical Milky Way color law of $\sim3$ results in an average brightness difference between the rungs of 0.012mag. Given that the $H_0$ difference is $\sim$0.20mag, this suggests very minimal impact on $H_0$ from differences in Milky Way $E(B-V)$ across rungs (See Table~\ref{tab:variants} for $\Delta H_0$). Lastly, it is important to note that small errors in the dust maps will result in different inferred SALT color parameter ($c$) of each light curve. Because the SALT color-luminosity relation $\beta\sim3$ is very similar to that Milky Way dust attenuation law $\sim3$, dust map differences are fortuitously `absorbed' by the Tripp estimator Eq.~\ref{eq:tripp}. 

Pantheon+ incorporate distance uncertainties in a covariance built from numerous systematic perturbations to the analysis, including perturbations characterizing the uncertainty in the underlying model of dust. The inferred $H_0$ for each of the systematic perturbations is shown in Figure 13 of Pantheon+ and shows a 0.2km/s/Mpc scatter in $H_0$ resulting from the uncertainty in the dust population parameters, which is well below the level of the $H_0$ tension. This is bolstered by the fact that because the Cepheids are only visible in edge-on late type/spiral galaxies in the second rung, and robust determinations of $H_0$ select for the similar types of host galaxies in the third rung of the distance ladder. While the underlying dust properties and differences between spiral galaxies and other galaxy types remain an open question, restricting the third rung of the distance ladder in SH0ES to only late type galaxies mitigates possible differences in galaxy properties between rungs. The SH0ES analysis explore an additional variant in which they allow all types of Hubble flow host galaxies (and therefore dust properties) and in this case find a $\Delta H_0=+0.3$, exacerbating the tension. 


The $\Delta H_0$ relative to the BS21 dust model is shown in Table~\ref{tab:variants} for the Guy et al. (2010) \cite{Guy2010} and Conley (2011) \cite{Conley2010} intrinsic scatter models. Instead of dust, the Guy et al. (2010) model describes SN Ia scatter with a grey luminosity scatter. While this model does not perform as well on the most recent Pantheon+ data (see Figure 5 of BS21), nonetheless while this model is agnostic to host-galaxy dust it only achieves a $\Delta H_0$ of -0.2km/s/Mpc. Likewise, Conley (2011) characterizes SN Ia scatter with empirical spectral variations and finds $\Delta H_0$ of -0.2km/s/Mpc.

\begin{table}
    \centering
    \begin{tabular}{r|c}
         Dust-Related Systematic & ~~$\Delta H_0$(km/s/Mpc) \\
        \hline 
\\
    \textbf{\textit{Cepheids}} & \\
          Milky Way (MW) Color Law from Fitzpatrick (1999), $R_V$ = 2.5 & +0.20 \\
          MW Color Law from Cardelli, Clayton, \& Mathis (1989), $R_V$ = 3.1 & +0.05 \\
          $R_W$ free global & +0.20 \\
           Intrinsic color subtracted \& MW Color Law from Fitzpatrick (1999) \& $R_V$ = 3.3 & +0.09 \\
           Intrinsic color subtracted \& MW Color Law from Fitzpatrick (1999) \& $R_V$ = free & +0.30 \\
          Intrinsic color subtracted \& $R_V$ (host mass-SFR) & +0.81 \\
          No dust treatment ($A_H$ values assumed to cancel) & +0.74 \\

    \\
   \textbf{\textit{SNe Ia}} & \\
         Milky Way (MW) Dust Map Absolute Scaling (4\%) & -0.02 \\
         MW Color Law from Cardelli, Clayton, \& Mathis (1989)  & +0.05   \\
         Grey Scatter Model from Conley (2010) &  -0.20  \\
         Spectral Scatter Model from Guy et al. (2011) & -0.22\\
         SN Ia/Host Dust Model Parameter Uncertainty 1 &  -0.02  \\
         SN Ia/Host Dust Model Parameter Uncertainty 2 &  -0.12  \\
         SN Ia/Host Dust Model Parameter Uncertainty 3 &  -0.13  \\
         All host galaxy morphologies/types in third rung&  +0.28  \\
        Hubble flow high mass hosts only $log_{10}(M_{\rm stellar}) > 10$  & -0.07 \\
    \end{tabular}
    \caption{Variants to the Pantheon+ and SH0ES analysis that assess the impact of dust. Negative $\Delta H_0$ is in the direction of reducing tension.}
    \label{tab:variants}
\end{table}

There have also been numerous independent crosschecks on the SNe Ia and Pantheon+ dust treatment. First, \cite{Burns_2018} derive SN Ia distances using infrared light curves thus reducing the impact of dust. They find reduced correlations with host-galaxy properties and SN Ia luminosity, they suggest that dust drives the correlations and their infrared observations explain the lack of correlation. Ultimately \cite{Burns_2018} report a value of the $H_0=73.2\pm2.3$ km/s/Mpc. 

Most recently, \cite{Dhawan_2023} assessed the impact of dust on a subset of the SNe Ia for which additional wavelengths from UV to NIR were available and with an improved SN Ia model were able to disentangle the dust and SN properties for 37 calibrator SNe and 67 Hubble flow SNe.  While only for a subset of the total Pantheon+ SNe, this importantly meant \cite{Dhawan_2023} could simultaneously compute an independent check on the Pantheon+ dust modeling and find $H_0=74.82\pm1.28$. 

A third major crosscheck has been made in \cite{Garnavich_2023} on the SH0ES methodology of restricting SN Ia hosts in the Hubble flow to match that of the second rung of Cepheid hosts. To avoid the SN Ia dust systematics imposed by restricting to late-type star-forming galaxies in both the second and third rungs of the distance ladder, \cite{Garnavich_2023} use what is colloquially called a 4-rung distance ladder. Using galaxy groups, \cite{Garnavich_2023} transfer the Cepheid calibration late-type galaxies to elliptical galaxies using the surface brightness fluctuations (SBF) technique of determining relative distances of early-type galaxies. These galaxies have different distributions of mass and dust and also host different populations of SNe Ia. While the introduction of an additional and less precise SBF rung of the distance ladder results in a larger uncertainty, the recovered central value is an important crosscheck on many host dust related systematics. \cite{Garnavich_2023} find $H_0=74.6\pm2.85$.

\subsection{Dust Impact on SN Ia constraints of Dark Energy and Dark Matter.}

While the above has shown that impact on inference of $H_0$ from Milky-Way and host-galaxy dust is minimal, the same cannot be said for inference of other cosmological parameters: Matter+Dark Matter Density ($\Omega_M$) and Equation of State of Dark Energy ($w$). The $H_0$ insensitivity is largely due to similarities between the second and third rung SNe Ia and hosts, but it is also attributable to the small redshift range in the Hubble flow (third) rung of the distance ladder $0.023<z<0.15$, which corresponds to a relatively small span of $<$ 1.5 billion years of cosmic history over which dust properties have little time to evolve. 

For inference of $\Omega_M$ and $w$, a much larger span of cosmic history is used to compile sufficient SN Ia statistics and the constraints of those models are also best performed at relatively higher redshifts. Samples of SNe Ia covering a very large span of cosmic history (currently 11 billion years) opens up the possibility for the evolution of dust parameters or the environments in which SNe Ia are found. Both \cite{bs21} and \cite{popovic21b} show that if uncertainties in dust populations of SN Ia hosts can evolve with redshift and this results in the largest systematic uncertainty in SN Ia cosmology today (on par with photometric calibration). Though, if correlations between dust properties and host-galaxy photometric properties (i.e. stellar mass, star formation rate, color) can be well constrained at low-redshift, then potential population evolution can be traced photometrically for future high redshift surveys. We note here that while there is much effort on survey calibration for future telescopes, there is comparatively less effort being placed on Milky Way and host-galaxy dust modeling, both of which are tractable issues with external datasets and effort. 

Future surveys such as The Vera Rubin Observatory Legacy Survey of Space and Time (LSST) and The Nancy Grace Roman Space Telescope (Roman) will be forced to tackle this challenge. LSST observes in optical wavelengths and will be more susceptible to the impact of host-galaxy dust and population evolution similarly to that of the Dark Energy Survey analysis. However, unlike the Dark Energy Survey, LSST will survey the entire southern sky (at much larger values of Milky Way $E(B-V)$) and will have to contend with increased systematics due to Milky Way dust. It will be important to improve uncertainties on Milky Way dust maps in regions of the sky where extinction is larger. Comparatively, Roman will observe in the infrared, which will make it less sensitive to dust for SNe Ia at moderate redshifts, though Roman will contend with the same dust population evolution systematics as other surveys when observing SNe Ia at redshift 3 (rest-frame optical).

\section{Conclusion}

In conclusion, we have explored the impact of dust on Cepheid and Type Ia Supernova distance indicators and the resulting impact on cosmological constraints, focusing on the measurement of the Hubble Constant ($H_0$). We show that the effects of dust propagate to changes in $H_0$ only if there are significant differences between different rungs of the distance ladder.
We discuss the impact of Milky-Way dust and host-galaxy dust and the associated uncertainties in these properties. In both cases, the differential measurements employed in the distance ladder approach for both Cepheids and SNe Ia largely cancel out the effects of dust, making them robust tools for measuring $H_0$. 

For Cepheids, which are observed in the Hubble Space Telescope (HST) Near Infrared (NIR) bands, the impact of dust is minimal. The use of Wesenheit magnitudes allows for the correction of dust effects based on directly observed quantities and the small value of the reddening ratio ($R_H$) in the NIR reduces the sensitivity to dust, which has been well-characterized for Milky Way-like extinction. For SNe Ia, we discuss the application of dust corrections to SNe Ia as a function of host-galaxy properties and as an important component of accounting for SN Ia intrinsic scatter. These dust-based corrections are applied consistently across rungs and uncertainties in the dust model parameters are characterized and accounted for and do not account for the observed Hubble tension.

Overall, the analysis presented in this paper reaffirms the robustness of current $H_0$ measurements and suggests that dust-related effects are not a significant source of tension in the determination of the Hubble Constant. However, we emphasize the importance of continued refinement in our understanding of dust properties and dust variation for cosmological studies beyond $H_0$ measurements.






\begin{acknowledgement}
Brout and Riess would like to thank Doug Finkbeiner for his useful comments on this chapter and discussions on the topic in general. 
This is a preprint of the following chapter: Dillon Brout and Adam Riess, The Impact of Dust on Cosmic Distance Indicators, published in The Hubble Constant Tension, edited by Eleonora DiValentino and Dillon Brout, expected 2024, Springer Nature reproduced with permission of Springer Nature Singapore Pte Ltd.. The final authenticated version will be available online at: http://dx.doi.org/[insert DOI when available]

\end{acknowledgement}






\bibliographystyle{unsrt}
\bibliography{chapter}

\end{document}